\documentclass[conference]{IEEEtran}
\IEEEoverridecommandlockouts
\usepackage{cite}
\usepackage{amsmath,amssymb,amsfonts}
\usepackage{algorithmic}
\usepackage{graphicx}
\usepackage{textcomp}
\usepackage{xcolor}
\usepackage{array}
\def\BibTeX{{\rm B\kern-.05em{\sc i\kern-.025em b}\kern-.08em
    T\kern-.1667em\lower.7ex\hbox{E}\kern-.125emX}}
\begin{document}

\title{A Foundation Model for Cross-Band CSI Reconstruction}

% \author{\IEEEauthorblockN{Hongpu Zhang$^{1}$, Shu Sun$^{1}$, Hangsong Yan$^{2}$, Jianhua Mo$^{1}$}
%     \IEEEauthorblockA{$^1$ School of Information Science and Electronic Engineering, Shanghai Jiao Tong University, Shanghai, China}
%     \IEEEauthorblockA{$^2$ Hangzhou Institute of Technology, Xidian University, Hangzhou, Zhejiang, China}
%     \IEEEauthorblockA{Corresponding author: Shu Sun (Email: shusun@sjtu.edu.cn)}
%     \thanks{This work is supported by the National Natural Science Foundation of China under Grants 62271310 and 62431014.}

\author{
\IEEEauthorblockN{Hongpu Zhang$^{1}$, Shu Sun$^{1}$, Ruifeng Gao$^{2}$, Tongjia Zhang$^{1}$ and Feng Yang$^{1}$}
    \IEEEauthorblockA{$^1$ School of Information Science and Electronic Engineering, Shanghai Jiao Tong University, Shanghai 200240, China}
    \IEEEauthorblockA{$^2$ School of Transportation and Civil Engineering, Nantong University, Nantong 226019, China}
    \IEEEauthorblockA{Corresponding author: Shu Sun (Email: shusun@sjtu.edu.cn)}
    % \thanks{This work is supported in part by the National Natural Science Foundation of China under Grants 62271310, 62571276, and 62431014, in part by the Science and Technology Commission Foundation of Shanghai under Grant 25DP1500100, and in part by the Natural Science Foundation of Nantong under Grant JC2023074.}
    }
\maketitle

\begin{abstract}
Acquiring dense high-frequency channel state information (CSI) is costly in
multi-band low-altitude wireless systems because pilot resources are limited and channel dimensions
change with the carrier frequency, bandwidth, and antenna array size. We address cross-band
CSI reconstruction, which recovers dense target-band CSI from dense source-band
CSI and sparse, noisy target-band pilots. We propose a foundation model that
represents every band in a common power--angle--delay spectrum and uses
radio-frequency metadata as auxiliary conditioning for an encoder--decoder.
The model uses pilot-guided cross-attention to fuse
source-band structure with target-band pilot, allowing one model to handle heterogeneous band pairs. It is trained by pilot-densification
pretraining followed by supervised cross-band fine-tuning. On ray-tracing, 3GPP,
and DeepMIMO datasets, the model lowers average normalized mean-square error
 by 6.1 dB over the state-of-the-art pair-specific baseline. It also transfers to two unseen pairs without paired fine-tuning,
achieving 7.5 and 7.1 dB gains over fully supervised baselines, and remains
effective when target pilots are noisy.
\end{abstract}


\section{Introduction}
\label{sec:intro}
Future low-altitude wireless networks are expected to operate in multiple frequency bands, including sub-6 GHz and millimeter-wave (mmWave) frequency bands.  This trend makes
accurate channel state information (CSI) essential, but also renders dense
high-frequency CSI expensive to acquire, especially for high frequency bands: the larger spatial--frequency dimensions
inflate pilot and feedback overhead, and the
higher-dimensional channels are sounded by proportionally fewer pilots under a fixed pilot budget, raising
sensitivity to pilot sparsity and noise~\cite{CEM}.
A natural way to reduce this cost is to reuse the channel of a lower-frequency,
more easily sounded source band. Out-of-band information has long been used this
way: to steer mmWave beam selection~\cite{oob} or to extrapolate the
channel across frequency and avoid explicit feedback~\cite{ELN}, and
learning-based methods more recently extend the idea to mapping channels across
antenna and frequency configurations~\cite{mapping}. The shared
premise is that bands sharing a propagation environment retain the same dominant
delay and angular structure, so a lower band carries a useful prior for
reconstructing the target-band CSI.

However, exploiting such information is difficult because the shared structure is
only partial: array responses, delay resolutions, path gains, and pilot
observations all vary with carrier frequency, bandwidth, and antenna
configuration~\cite{oob,marine}. A source-band CSI matrix therefore cannot be mapped directly
to the target band, and sparse target-band pilots are still needed to anchor the
reconstruction. Classical interpolation uses only pilots and ignores the
cross-band prior, while most learning-based CSI work targets feedback
compression~\cite{csinet}, which does not address reconstruction from sparse
cross-band observations and faces persistent generalization and
configuration-dependence challenges~\cite{CSIFeed}. 

These requirements motivate a foundation-model approach: a single model should learn reusable channel representations and
remain aware of varying radio-frequency (RF) configurations. Recent wireless foundation models~\cite{eng}
pursue this direction through large-scale self-supervised pretraining: For instance, LWM learns
task-agnostic channel embeddings~\cite{LWM}, WiFo targets one-for-all
channel prediction across heterogeneous configurations~\cite{WiFo}, and
ContraWiMAE learns masked--contrastive channel
representations~\cite{MAL}. These works show the promise of pretrained,
zero-shot channel representations, but mainly target general representation or
prediction. Cross-band reconstruction is a related yet less explored setting,
where a dense source-band channel and a few noisy target-band pilots are jointly
available for target-band CSI recovery.

In this work, we propose an encoder--decoder for
cross-band CSI reconstruction with source-band CSI and target-band
pilots as inputs. The proposed method combines a common delay--angle
representation, pilot-guided source--target fusion, and RF metadata
conditioning in one network. Our main contributions are as follows.
\begin{itemize}
\item \textbf{A foundation model for heterogeneous cross-band reconstruction.} We
place heterogeneous CSI on a common delay--angle grid and fuse source CSI with
sparse target pilots in one backbone, while RF prompts provide auxiliary band
descriptors for heterogeneous configurations.
\item \textbf{A task-aligned two-stage training scheme.} A pilot-densification
pretext is followed by cross-band fine-tuning, first learning to densify sparse
self-pilots and then adapting the same architecture to source--target fusion.
\item \textbf{Extensive experimental validation.} Across six band pairs, the proposed
method lowers normalized mean-square error (NMSE) by $6.1$~dB on average over the state-of-the-art full-shot baseline,
transfers to unseen band pairs zero-shot, and remains robust under noisy pilots.
\end{itemize}

\section{System Model and Problem Formulation}

\label{sec:system}

\subsection{Channel Model}

We consider a base station (BS) operating over a set of
carrier bands $\mathcal{B}$. For each band $b\in\mathcal{B}$, the BS is equipped
with a band-specific uniform planar array (UPA) and orthogonal frequency division multiplexing (OFDM) numerology. These UPAs share the same BS site
but may have different array dimensions and carrier-dependent element spacings.
Specifically, the UPA for band $b$ has $N_y^{(b)} \times N_z^{(b)}$ elements,
with $N_t^{(b)}=N_y^{(b)}N_z^{(b)}$, and its OFDM grid has $N_c^{(b)}$
subcarriers with spacing $\Delta f^{(b)}$. The baseband frequency of subcarrier $n$
is $f_n^{(b)}=(n-\lfloor N_c^{(b)}/2\rfloor)\Delta f^{(b)} + f^{(n)}_c$, the center wavelength is
$\lambda_c^{(b)}=c_0/f_c^{(b)}$ with $c_0$ the speed of light, and the UPA takes
half-wavelength spacing $d_y^{(b)}=d_z^{(b)}=\lambda_c^{(b)}/2$. A path $\ell$ has
delay $\tau_\ell$ and direction cosines
$u_\ell=\sin\theta_\ell\cos\phi_\ell$, $v_\ell=\sin\theta_\ell\sin\phi_\ell$,
collected as $\mathbf{o}_\ell=(u_\ell,v_\ell)$. The subcarrier channel and its
steering vector are
\begin{align}
    \mathbf{h}_n^{(b)}
    &=
    \sum_{\ell=1}^{L}
    \alpha_\ell^{(b)}
    e^{-\mathrm{j}2\pi f_{n}^{(b)}\tau_\ell}
    \mathbf{a}_n^{(b)}(\mathbf{o}_\ell), \\
    [\mathbf{a}_n^{(b)}(\mathbf{o}_\ell)]_{m_y,m_z}
    &=
    \tfrac{1}{\sqrt{N_t^{(b)}}}
    \exp\!\Big(\!-\mathrm{j}2\pi
    \tfrac{(m_y d_y^{(b)}u_\ell+m_z d_z^{(b)}v_\ell)f_{n}^{(b)}}{c_0}\Big),
\end{align}
with $m_y=0,\ldots,N_y^{(b)}-1$ and $m_z=0,\ldots,N_z^{(b)}-1$. 
The UPA response is vectorized to form
$\mathbf{h}_n^{(b)}\in\mathbb{C}^{N_t^{(b)}}$, and stacking all subcarriers gives
$\mathbf{H}^{(b)} \in \mathbb{C}^{N_c^{(b)} \times N_t^{(b)}}$. Co-located bands
share almost similar path geometry
$\{(\tau_\ell,\mathbf{o}_\ell)\}$, while the path gains
$\alpha_\ell^{(b)}$, array dimensions, and RF parameters are band-dependent.

The target band is observed only through sparse, noisy pilots. Under a uniform
lattice pattern with strides $(\kappa_f, \kappa_a)$,
\begin{equation}
    \bar{\mathbf{H}}_{pilot}
    =
    \mathbf{M}_{\kappa_f,\kappa_a} \odot (\bar{\mathbf{H}}_t + \mathbf{N}),
\end{equation}
where $\bar{\mathbf{H}}_t$ is the normalized target CSI, $\mathbf{N}$ is pilot
noise at a prescribed signal-to-noise ratio (SNR), and the binary mask $\mathbf{M}_{\kappa_f,\kappa_a}$
retains every $\kappa_f$-th subcarrier and $\kappa_a$-th antenna (entry $1$ when
$n\bmod\kappa_f=0$ and $m\bmod\kappa_a=0$, else $0$), giving pilot density
$1/(\kappa_f\kappa_a)$. Each band has RF configuration
$(f_c^{(b)},N_c^{(b)},\Delta f^{(b)},N_t^{(b)})$, with network prompt
$\mathbf{r}^{(b)}$ specified in Section~\ref{sec:method}.

\subsection{Problem Formulation}

Co-located bands that share a propagation environment retain the same
geometry-induced delay--angle structure while differing in carrier frequency,
array size, subcarrier grid, and per-path gains, as mentioned in
Section~\ref{sec:intro}. We exploit this property in the following
setting: an easily sounded \emph{source} band $s$ assists the acquisition of a
costly \emph{target} band $t$, and our goal is to recover the dense target CSI
from the source observation together with only a few target-band pilots.

Formally, given the normalized source CSI $\bar{\mathbf{H}}_s$, the sparse noisy
target pilots $\bar{\mathbf{H}}_{pilot}$, and the RF prompts
$\mathbf{r}_s, \mathbf{r}_t$, we estimate the dense target CSI
$\bar{\mathbf{H}}_t$. We learn a mapping $\mathcal{F}_{\Theta}$
that is shared across all pairs,
\begin{equation}
    \widehat{\bar{\mathbf{H}}}_t
    =
    \mathcal{F}_{\Theta}\!\big(
    \bar{\mathbf{H}}_s,\, \bar{\mathbf{H}}_{pilot},\,
    \mathbf{r}_s,\, \mathbf{r}_t \big),
\end{equation}
and optimize its parameters $\Theta$ by minimizing the expected target CSI
reconstruction error, with the training objective
detailed in Section~\ref{sec:method}.

\section{Proposed Method}
\label{sec:method}

\begin{figure*}[!t]
\centering
\includegraphics[width=0.90\textwidth]{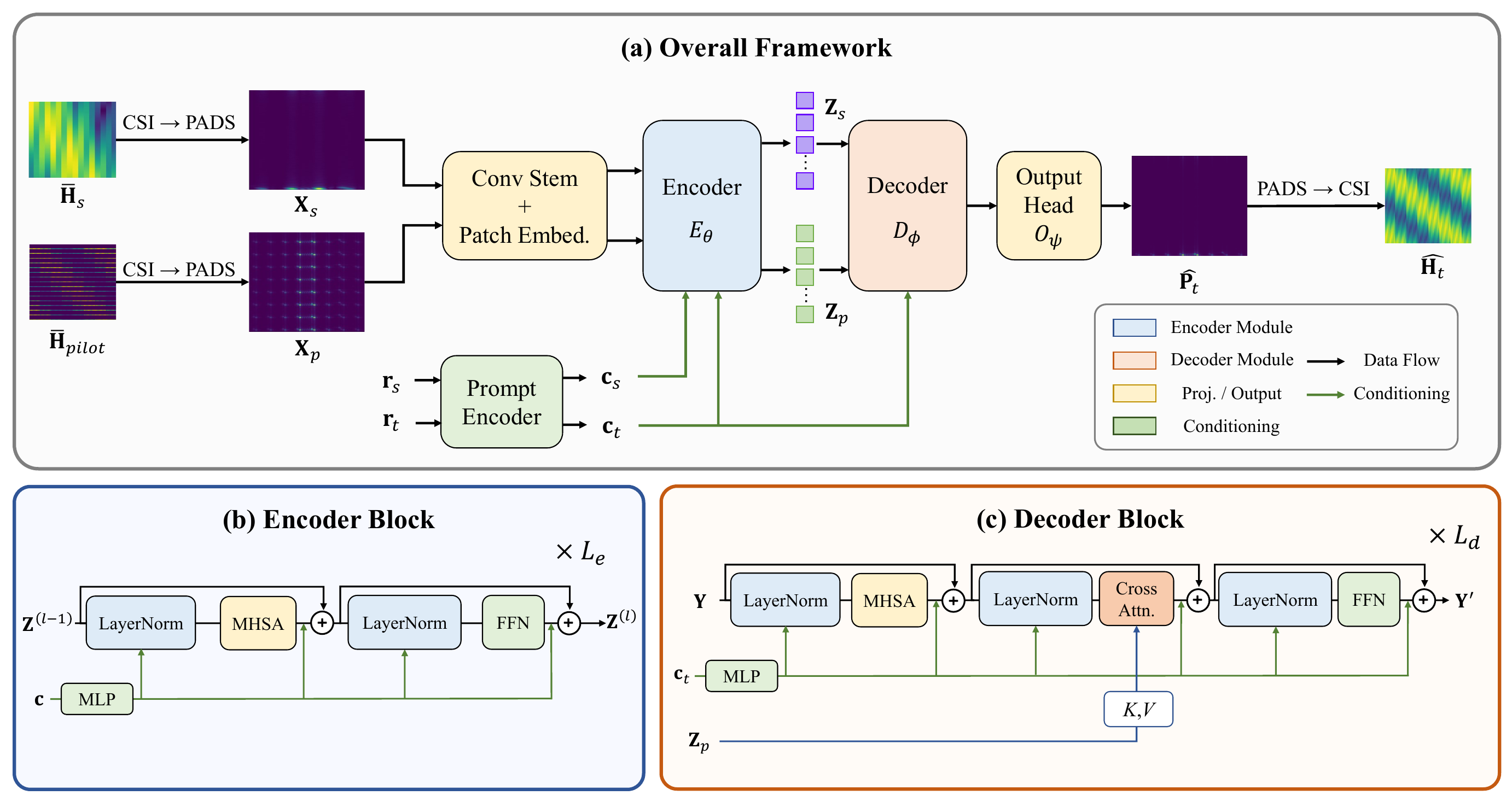}
\caption{Model architecture of the proposed method, including the overall framework and the
encoder/decoder block designs.}
\label{fig:model_architecture}
\end{figure*}
In this section, we present the proposed encoder--decoder
network for cross-band CSI reconstruction. The model architecture, including
the overall data flow and the encoder/decoder block designs, is shown in
Fig.~\ref{fig:model_architecture}.

\subsection{Unified Power-Angle-Delay Spectrum Representation}

To place the band-dependent CSI matrices $\mathbf{H}^{(b)}$ on a common grid, we map each CSI matrix to a fixed-size power-angle-delay spectrum (PADS). For each band, we build a delay
dictionary $\mathbf{W}_d^{(b)} \in \mathbb{C}^{H \times N_c^{(b)}}$ and an
angular dictionary $\mathbf{W}_a^{(b)} \in \mathbb{C}^{N_t^{(b)} \times W}$.Let $T_{\max}^{(b)}=1/\Delta f^{(b)}$,
$\tau_i=iT_{\max}^{(b)}/H$, $u_j=2j/W-1$, and
$q_m=m-(N_t^{(b)}-1)/2$. The dictionaries are constructed as
\begin{align}
    [\mathbf{W}_d^{(b)}]_{i,n}
    &=
    H^{-1/2}
    \exp\!\left(\mathrm{j}2\pi \tau_i f_n^{(b)}\right), \\
    [\mathbf{W}_a^{(b)}]_{m,j}
    &=
    (N_t^{(b)})^{-1/2}
    \exp\!\left(-\mathrm{j}\pi q_m u_j\right),
\end{align}
Given the dataset-level CSI and PADS scales
$s_H^{(b)}$ and $s_P^{(b)}$, the model input is
\begin{align}
    \mathbf{P}^{(b)}
    &=
    \mathbf{W}_d^{(b)}
    (\mathbf{H}^{(b)}/s_H^{(b)})
    \mathbf{W}_a^{(b)}, \\
    \mathbf{X}^{(b)}
    &=
    [\Re(\mathbf{P}^{(b)}/s_P^{(b)}),
    \Im(\mathbf{P}^{(b)}/s_P^{(b)})]
    \in\mathbb{R}^{2\times H\times W}.
\end{align}
The scale statistics are computed from the datasets as described in
Section~\ref{sec:exp_setup}.
This gives all bands a shared delay-angle input grid while preserving
propagation sparsity.

\subsection{Band-Aware Encoder}
We first tokenize the input. A convolutional stem downsamples the PADS
$\mathbf{X}\in\mathbb{R}^{2\times H\times W}$ by $4\times$, and a patch embedding
with patch size $p_h \times p_w$ projects the result into $N=HW/p_hp_w$ tokens of dimension $D$. We prepend a learnable class token
$\mathbf{t}_{cls}\in\mathbb{R}^{D}$ and add a fixed two-dimensional
sinusoidal positional encoding $\mathbf{E}_{pos}\in\mathbb{R}^{(N+1)\times D}$,
yielding the token sequence $\mathbf{Z}^{0}\in\mathbb{R}^{(N+1)\times D}$,
\begin{equation}
    \mathbf{Z}^{0}
    =
    \big[\mathbf{t}_{cls};\,
    \mathrm{PatchEmbed}(\mathrm{Stem}(\mathbf{X}))\big]
    +\mathbf{E}_{pos},
\end{equation}
where, for a patch at grid position $(p,q)$, the positional encoding
concatenates one-dimensional sinusoidal codes of its row and column,
$[\mathbf{E}_{pos}]_{p,q}=[\mathbf{e}(p);\mathbf{e}(q)]$, with
$\mathbf{e}(x)\in\mathbb{R}^{D/2}$ defined by $e_{2i}(x)=\sin(x\,\omega_i)$,
$e_{2i+1}(x)=\cos(x\,\omega_i)$ and $\omega_i=10000^{-4i/D}$.

To make these tokens band-aware, we summarize the RF configuration of band $b$
as the prompt $\mathbf{r}^{(b)}\in\mathbb{R}^{4}$,
\begin{equation}
    \label{eq:prompt}
    \mathbf{r}^{(b)}
    =
    [\log f_c^{(b)}, \log N_c^{(b)}, \log \Delta f^{(b)}, \log N_t^{(b)}],
\end{equation}
and lift it by random Fourier features and a two-layer multilayer perceptron (MLP) into a conditioning
vector $\mathbf{c}^{(b)}\in\mathbb{R}^{D}$. Rather than appending
$\mathbf{c}^{(b)}$ as an extra token, every Transformer block injects it through
adaptive layer normalization (AdaLN)~\cite{dit},
\begin{equation}
    \mathrm{AdaLN}(\mathbf{Y};\mathbf{c})
    =
    \big(1+\mathbf{g}(\mathbf{c})\big)
    \odot
    \mathrm{LayerNorm}(\mathbf{Y})
    +
    \mathbf{b}(\mathbf{c}),
\end{equation}
where the scale $\mathbf{g}(\mathbf{c})\in\mathbb{R}^{D}$ and shift
$\mathbf{b}(\mathbf{c})\in\mathbb{R}^{D}$ are linear functions of the RF
condition. The carrier frequency, bandwidth, and array size therefore reshape
the feature statistics for each sub-layer.

Each of the $L_e$ encoder blocks adopts the pre-norm form, applying
AdaLN-modulated multi-head self-attention (MHSA)~\cite{attn} and then a SwiGLU feed-forward
network (FFN), each wrapped in a gated residual connection. For block $\ell$ with
input $\mathbf{Z}^{\ell-1}$,
\begin{align}
    \mathbf{Z}'
    &=
    \mathbf{Z}^{\ell-1}
    + \mathbf{s}_a \odot
    \mathrm{MHSA}\!\big(\mathrm{AdaLN}(\mathbf{Z}^{\ell-1};\mathbf{c})\big), \\
    \mathbf{Z}^{\ell}
    &=
    \mathbf{Z}'
    + \mathbf{s}_f \odot
    \mathrm{FFN}\!\big(\mathrm{AdaLN}(\mathbf{Z}';\mathbf{c})\big),
\end{align}
where the residual gates $\mathbf{s}_a,\mathbf{s}_f\in\mathbb{R}^{D}$, together
with the scale and shift of the two AdaLN instances, are all produced from
$\mathbf{c}$ by a linear layer. The encoder output is the final-block
sequence $\mathbf{Z}=\mathbf{Z}^{L_e}$.

The encoder $E_{\theta}$ handles both inputs: it processes the full source PADS
$\mathbf{X}_s$ under the source condition and the sparse target-pilot PADS
$\mathbf{X}_p$ under the target condition,
\begin{equation}
    \mathbf{Z}_s = E_{\theta}(\mathbf{X}_s,\mathbf{c}_s),\qquad
    \mathbf{Z}_p = E_{\theta}(\mathbf{X}_p,\mathbf{c}_t),
\end{equation}
with $\mathbf{Z}_s,\mathbf{Z}_p\in\mathbb{R}^{(N+1)\times D}$. Weight sharing
places the source and pilot features in a common token space that the decoder
can fuse directly, while the prompt specializes the same weights to each band.

\subsection{Pilot-Guided Decoder}

The decoder starts from the encoded source sequence
$\mathbf{Y}^{0}=\mathbf{Z}_s$ and uses the target-pilot sequence
$\mathbf{Z}_p$ as a cross-attention context. Since the output belongs to
the target band, every decoder block is conditioned on $\mathbf{c}_t$.

Each of the $L_d$ decoder blocks has three sub-layers,
all conditioned on $\mathbf{c}_t$: MHSA, cross-attention, and a SwiGLU FFN. The
MHSA and FFN take the same form as in the encoder. The cross-attention module draws queries from the current source
stream $\mathbf{Y}$ and keys/values from the pilots $\mathbf{Z}_p$,
\begin{equation}
    \mathbf{Y}
    \leftarrow
    \mathbf{Y}
    + \mathbf{s}_x \odot
    \mathrm{CrossAttn}\!\big(\mathrm{AdaLN}(\mathbf{Y};\mathbf{c}_t),\,\mathbf{Z}_p\big),
\end{equation}
where, for a single head,
\begin{equation}
\mathrm{CrossAttn}(\mathbf{Y},\mathbf{Z}_p)=
\mathrm{Softmax}\!\big(\mathbf{Y}\mathbf{W}_Q\mathbf{W}_K^{\top}\mathbf{Z}_p^{\top}/\sqrt{d}\big)
\mathbf{Z}_p\mathbf{W}_V
\end{equation}
The source stream thus queries
target-band evidence while the pilot representation stays fixed. After the final block, an output head
$O_{\psi}$ reshapes the $N$ patch tokens $\mathbf{Y}^{L_d}_{1:N}$ and upsamples it to the dense target PADS
$\widehat{\mathbf{P}}_t\in\mathbb{R}^{2\times H\times W}$, which the fixed
pseudo-inverse dictionaries map back to CSI:
\begin{equation}
    \widehat{\mathbf{H}}_t
    =
    (\mathbf{W}_d^{(t)})^{\dagger}\,
    \widehat{\mathbf{P}}_t\,
    (\mathbf{W}_a^{(t)})^{\dagger}.
\end{equation}

\subsection{Two-Stage Training}

Both stages draw pilots from the uniform lattice of Section~\ref{sec:system}
and they reuse the same shared encoder $E_{\theta}$, decoder $D_{\phi}$, and output head $O_{\psi}$, differing only in how these modules are fed. All parameters are updated in both stages.

The first stage is \emph{pilot-densification pretraining}.
Given a dense PADS sample $\mathbf{X}^{(b)}$, we sample a pilot lattice from its
CSI at a randomly chosen density, add Gaussian noise, and transform the sparse
observation back to PADS as $\mathbf{X}_{pilot}^{(b)}$. The decoder receives the same tokens as both query and context,
\begin{equation}
    \widehat{\mathbf{X}}^{(b)}
    =
    O_{\psi}\big(D_{\phi}(\mathbf{Z}^{(b)},\mathbf{Z}^{(b)},\mathbf{c}^{(b)})\big),
    \mathbf{Z}^{(b)}=E_{\theta}(\mathbf{X}_{pilot}^{(b)},\mathbf{c}^{(b)}).
\end{equation}
The target is the dense PADS $\mathbf{X}^{(b)}$, and the loss is its PADS-domain
normalized mean square error (NMSE),
\begin{equation}
    \mathcal{L}_{pt}
    =
    \sum_i\|\widehat{\mathbf{X}}_i-\mathbf{X}_i\|_F^2
    \,\big/\,
    \sum_i\|\mathbf{X}_i\|_F^2 .
\end{equation}

\begin{table*}[!t]
\caption{Dataset composition, samples per band, and CSI configuration.}
\label{tab:dataset_composition}
\centering
\scriptsize
\setlength{\tabcolsep}{2.8pt}
\renewcommand{\arraystretch}{1}

\begin{tabular}{@{}
>{\centering\arraybackslash}m{0.08\textwidth}
>{\centering\arraybackslash}m{0.28\textwidth}
>{\centering\arraybackslash}m{0.08\textwidth}
>{\centering\arraybackslash}m{0.08\textwidth}
>{\centering\arraybackslash}m{0.05\textwidth}
>{\centering\arraybackslash}m{0.05\textwidth}
>{\centering\arraybackslash}m{0.09\textwidth}
>{\centering\arraybackslash}m{0.07\textwidth}
>{\centering\arraybackslash}m{0.10\textwidth}
@{}}
\hline
\textbf{Source} & \textbf{Dataset scenes} & \textbf{Band (GHz)} & \textbf{Samples} & \textbf{$N_c$} & \textbf{$N_t$} & \textbf{$\Delta f$ (kHz)} & \textbf{UPA} & \textbf{Test scene} \\
\hline

&
& 2.4 & 203K & 64 & 16 & 60 & $4{\times}4$ & \\
\cline{3-8}
&
& 3.5 & 203K & 64 & 16 & 90 & $4{\times}4$ & \\
\cline{3-8}
RT
& BME, D4, EMB, LA
& 15 & 203K & 128 & 32 & 90 & $8{\times}4$ & EMB \\
\cline{3-8}
&
& 28 & 203K & 128 & 64 & 180 & $8{\times}8$ & \\
\cline{3-8}
&
& 40 & 203K & 128 & 64 & 180 & $8{\times}8$ & \\
\hline

&
& 2.4 & 300K & 64 & 16 & 60 & $4{\times}4$ & \\
\cline{3-8}
3GPP
& UMa, RMa, InH, InF, UMi
& 3.5 & 300K & 64 & 16 & 90 & $4{\times}4$ & InH \\
\cline{3-8}
&
& 15 & 300K & 128 & 32 & 90 & $8{\times}4$ & \\
\cline{3-8}
&
& 28 & 300K & 128 & 64 & 180 & $8{\times}8$ & \\
\hline

DeepMIMO
& city 0--19 except 4; ASU campus at 3.5 GHz
& 3.5 & 1,385K & 64 & 16 & 90 & $4{\times}4$ & city 0/5/15 \\
\cline{3-8}
&
& 28 & 1,303K & 128 & 64 & 180 & $8{\times}8$ & \\
\hline
\end{tabular}
\end{table*}

The second stage is \emph{cross-band fine-tuning}. For each paired sample, the source band provides the dense PADS
$\mathbf{X}_s$, while the target band provides the sparse target-pilot PADS
$\mathbf{X}_p$ and the dense CSI supervision,
\begin{equation}
    \widehat{\mathbf{X}}_t
    =
    O_{\psi}\big(
    D_{\phi}(E_{\theta}(\mathbf{X}_s,\mathbf{c}_s),\,
    E_{\theta}(\mathbf{X}_p,\mathbf{c}_t),\,\mathbf{c}_t)\big),
\end{equation}
Then $\widehat{\mathbf{X}}_t$ is
mapped to CSI by the inverse dictionaries and the loss
$\mathcal{L}_{ft}$ is the same NMSE evaluated in the CSI domain, between
$\widehat{\mathbf{H}}_t$ and $\bar{\mathbf{H}}_t$. For multi-pair training, each
batch contains a single frequency pair.

\section{Experimental Evaluation}
\subsection{Experimental Setup}
\label{sec:exp_setup}
\textbf{Datasets and splits.}
We evaluate the proposed method on CSI datasets from ray-tracing (RT)\footnote{https://github.com/Zetahp/RT-Datasets}, 3GPP stochastic by QuaDRiGa~\cite{quad}, and
DeepMIMO channels~\cite{deepmimo}, with scene lists, sample counts, and CSI
configurations summarized in Table~\ref{tab:dataset_composition}. In the RT
subset, Biomedical Engineering Buildings (BME), Dong 4 Area (D4), and East Middle Buildings (EMB) are derived from Shanghai Jiao Tong University scenes
whose maps are extracted from OpenStreetMap~\cite{osm}, Los Angeles (LA) is self-built, and all RT CSI
samples are generated by Sionna RT~\cite{sionna}. All the CSI samples correspond to far-field channels~\cite{Far}. The two stages use separate pools: pretraining uses single-band CSI from 27 scenes (EMB and DeepMIMO city~0 held out for validation); fine-tuning uses six frequency pairs from 22 scenes, and testing uses five held-out scenes. Per-scene, per-band root mean square (RMS) scales for complex CSI and for the real/imaginary PADS are computed once and reused across training, validation, and testing.

\textbf{Task and evaluation.}
Each test sample contains dense source-band CSI $\mathbf{H}_s$ and noisy sparse
target-band pilots $\bar{\mathbf{H}}_{pilot}$; the output is dense
target-band CSI $\widehat{\mathbf{H}}_t$. Fine-tuning uses six
supervised frequency pairs:
2.4$\rightarrow$15, 2.4$\rightarrow$28, 3.5$\rightarrow$15,
3.5$\rightarrow$28, 15$\rightarrow$28, and 15$\rightarrow$40 GHz. Pair-specific
zero-shot settings
are given per experiment. Unless otherwise noted, target pilots use a uniform lattice with
frequency stride 8 and antenna stride 1 (12.5\% density), and pilot noise is
10--20 dB SNR in training. We report CSI-domain NMSE at 15 dB test SNR, with all
methods on the same test subset and results averaged over each pair's test scenes.

\begin{table}[!t]
\caption{Network Configuration.}
\label{tab:model_config}
\centering
\scriptsize
\setlength{\tabcolsep}{4pt}
\renewcommand{\arraystretch}{1}

\begin{tabular}{@{}
>{\centering\arraybackslash}m{0.34\linewidth}
>{\centering\arraybackslash}m{0.52\linewidth}
@{}}
\hline
\textbf{Item} & \textbf{Value} \\
\hline
PADS size $2\times H\times W$ & $2{\times}256{\times}256$ \\
RF prompt dim. & 4 \\
Embed. dim. $D$ & 384 \\
Enc. / Dec. depth $L_e / L_d$ & 6 / 6 \\
Attn. heads & 6 \\
Stem downsampling & $4{\times}$ \\
Patch size $p_h \times p_w$ & $8 \times 8$ \\
Params. & 40.7M \\
\hline
\end{tabular}
\end{table}

\begin{table}[!t]
\caption{Training configuration.}
\label{tab:train_hparams}
\centering
\scriptsize
\setlength{\tabcolsep}{4pt}
\renewcommand{\arraystretch}{1}

\begin{tabular}{@{}
>{\centering\arraybackslash}m{0.30\linewidth}
>{\centering\arraybackslash}m{0.28\linewidth}
>{\centering\arraybackslash}m{0.28\linewidth}
@{}}
\hline
\textbf{Item} & \textbf{Pretraining} & \textbf{Fine-tuning} \\
\hline
Loss & PADS-domain NMSE & CSI-domain NMSE \\
Epochs & 20 & 150 \\
Batch size & 512 & 256 \\
Learning rate & $2{\times}10^{-4}$ & $1.0{\times}10^{-4}$ \\
Weight decay & 0.05 & 0.01 \\
Warmup steps & 2000 & 500 \\
Pilot stride $(\kappa_f,\kappa_a)$ & $\{1,4,8,16,32\}{\times}1$ & $8{\times}1$ \\
Pilot SNR & 10--20 dB & 10--20 dB \\
\hline
\end{tabular}
\end{table}

\textbf{Baselines.}
All baselines output the dense target CSI $\widehat{\mathbf{H}}_t$. \textbf{Linear
interpolation} and \textbf{cubic spline interpolation} are non-learned pilot-only methods:
their inputs are the noisy sparse target pilots
$\bar{\mathbf{H}}_{pilot}$ and the pilot mask, and they independently
interpolate missing subcarriers for each antenna. \textbf{U-Net}~\cite{unet}
is a learned pilot-only
network with a three-level encoder--decoder and is
trained to predict the target CSI. \textbf{CsiNet}~\cite{csinet} is adapted
to cross-band reconstruction by using the source CSI $\mathbf{H}_s$ and target pilots as two input branches: a CsiNet-style source encoder and a target-pilot encoder produce
latent vectors, which are then fused and decoded. \textbf{LWM}~\cite{LWM} uses source CSI and target pilots
converted to patch tokens, and the LWM backbone is fine-tuned together with a cross-attention head to recover the target CSI. The proposed method uses the same
source CSI and target pilots as the source-aware baselines, with RF prompts. Learned baselines use the same pilot mask and SNR, and target supervision. They are optimized with AdamW under CSI-domain NMSE, with checkpoints selected on the validation split. To give the baselines a favorable setting, we allow target-scene supervised
training: for each cross-band test scene, samples are split $70/10/20$ into
train/validation/test. A separate baseline is trained on the $70\%$ training split for each frequency pair and scene, optimized with AdamW under CSI-domain NMSE. All methods are evaluated on the same $20\%$ test subset.

\subsection{Implementation Details}

All experiments use the network configuration in
Table~\ref{tab:model_config}, whose 40.7M parameters cover the tokenizer, RF
prompt encoder, Transformer encoder, decoder, and output head. Table~\ref{tab:train_hparams} lists the two-stage training hyperparameters,
optimized with AdamW under cosine learning-rate decay, gradient clipping at 1.0,
and fp32 precision. The fine-tuning ranges reflect per-pair selection on the
validation split. Experiments run on a server with two AMD EPYC 9454 48-Core CPUs and
four NVIDIA GeForce RTX 4090 GPUs.

\subsection{Numerical Results}

We first evaluate cross-band CSI reconstruction, and then examine
pilot-noise robustness, zero-shot frequency-pair transfer, ablation study, and inference cost.

Table~\ref{tab:main_results} compares all methods on six supervised frequency pairs at a 15~dB pilot SNR. Our method achieves the best NMSE on every pair, improving over the most competitive baseline by $6.1$~dB on average. The gains are smaller when the full-shot U-Net is already strong, but consistently show that source CSI provides complementary structure beyond target pilots alone. The results also highlight the need for both cross-band information and RF-conditioned fusion. Linear interpolation is competitive on easier 15~GHz targets but degrades as bandwidth and array size grow, and U-Net cannot exploit source-band CSI. CsiNet and LWM use source CSI, yet their inconsistent gains show that cross-band information must be fused with band awareness rather than treated as generic input.

\begin{table}[!t]
\caption{Main comparison across frequency pairs.}
\label{tab:main_results}
\centering
\scriptsize
\setlength{\tabcolsep}{2.2pt}
\renewcommand{\arraystretch}{1}

\begin{tabular}{@{}
>{\centering\arraybackslash}m{0.18\linewidth}
>{\centering\arraybackslash}m{0.12\linewidth}
>{\centering\arraybackslash}m{0.12\linewidth}
>{\centering\arraybackslash}m{0.12\linewidth}
>{\centering\arraybackslash}m{0.12\linewidth}
>{\centering\arraybackslash}m{0.12\linewidth}
>{\centering\arraybackslash}m{0.12\linewidth}
@{}}
\hline
\textbf{Method} 
& \textbf{$2.4{\to}15$} 
& \textbf{$2.4{\to}28$} 
& \textbf{$3.5{\to}15$} 
& \textbf{$3.5{\to}28$} 
& \textbf{$15{\to}28$} 
& \textbf{$15{\to}40$} \\
\hline
Linear & -21.79 & -16.14 & -21.79 & -15.71 & -16.14 & -9.38 \\
Spline & -19.61 & -11.66 & -19.61 & -10.31 & -11.66 & -4.26 \\
CsiNet & -9.85 & -5.97 & -9.85 & -6.89 & -6.58 & -1.06 \\
U-Net & -20.93 & -21.24 & -20.77 & -19.23 & -21.29 & -19.16 \\
LWM & -18.85 & -11.53 & -17.61 & -9.22 & -11.31 & -9.43 \\
Proposed Method & \textbf{-26.93} & \textbf{-25.34} & \textbf{-26.98} & \textbf{-28.98} & \textbf{-25.58} & \textbf{-27.03} \\
\hline
\end{tabular}
\end{table}

\begin{figure}[!t]
\centering
\includegraphics[width=0.66\columnwidth]{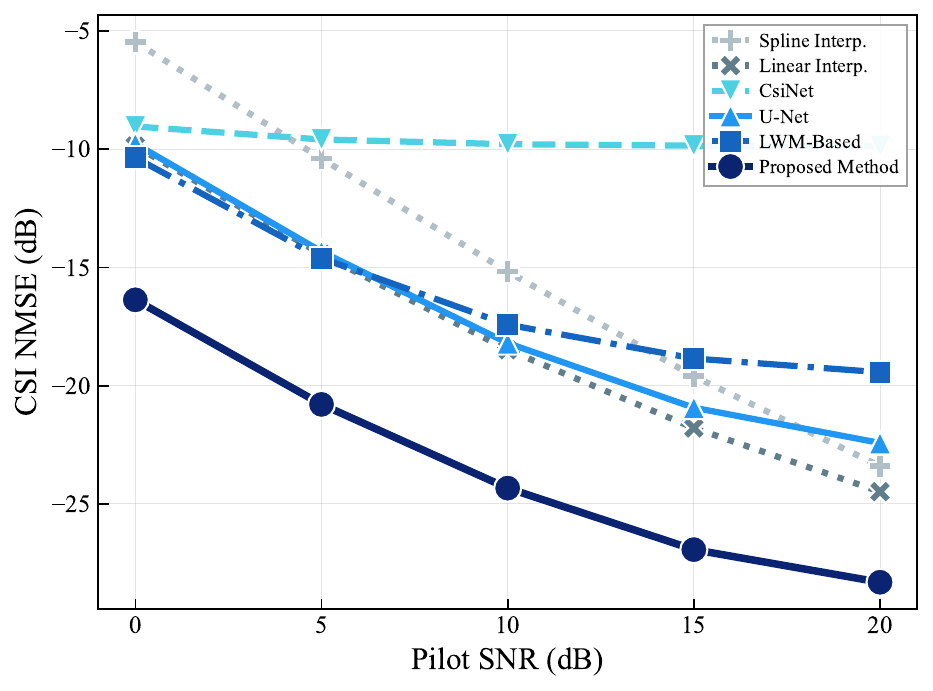}\\[-0.8ex]
\small (a) $2.4\rightarrow15$ GHz\\[0.8ex]
\includegraphics[width=0.66\columnwidth]{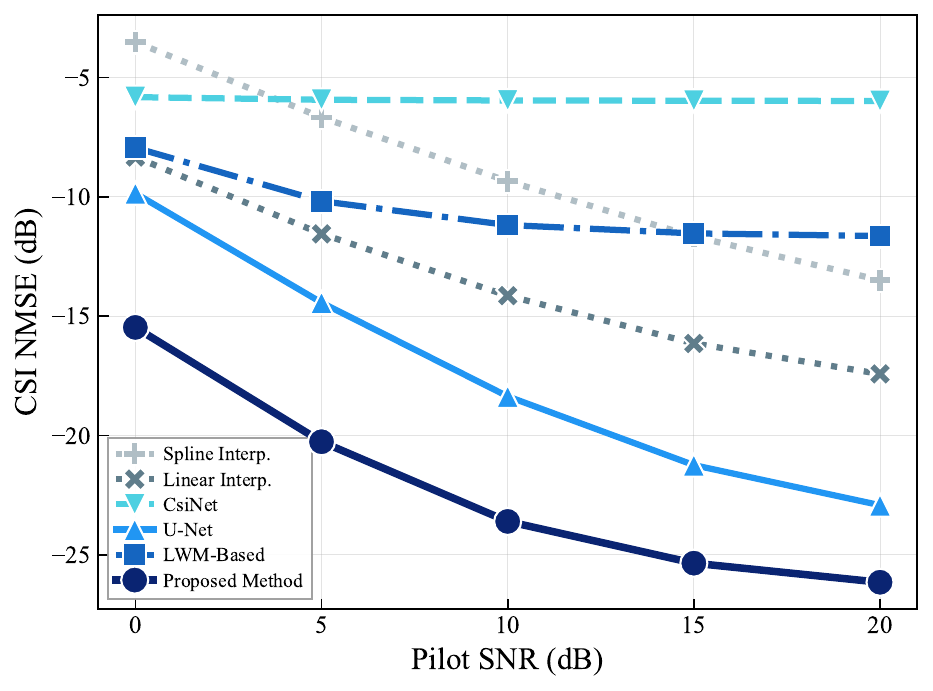}\\[-0.8ex]
\small (b) $2.4\rightarrow28$ GHz
\caption{Robustness of cross-band reconstruction to pilot noise.}
\label{fig:snr_results}
\end{figure}

Fig.~\ref{fig:snr_results} measures pilot-noise robustness by sweeping the
target-pilot SNR from 0 to 20~dB on $2.4\rightarrow15$ and
$2.4\rightarrow28$ GHz, while keeping the model parameters, pilot mask, and test
subset fixed. The proposed method stays best across the range. At low SNRs all methods are limited by noise; at high SNRs the error is dominated by missing target samples, so interpolation and
pilot-only networks approach a pilot-density ceiling, while the gap our method
retains shows that cross-band geometry adds resolution beyond what
sparse pilots provide, especially for the higher-dimensional
28~GHz target.

\begin{figure}[!t]
\centering
\includegraphics[width=0.66\columnwidth]{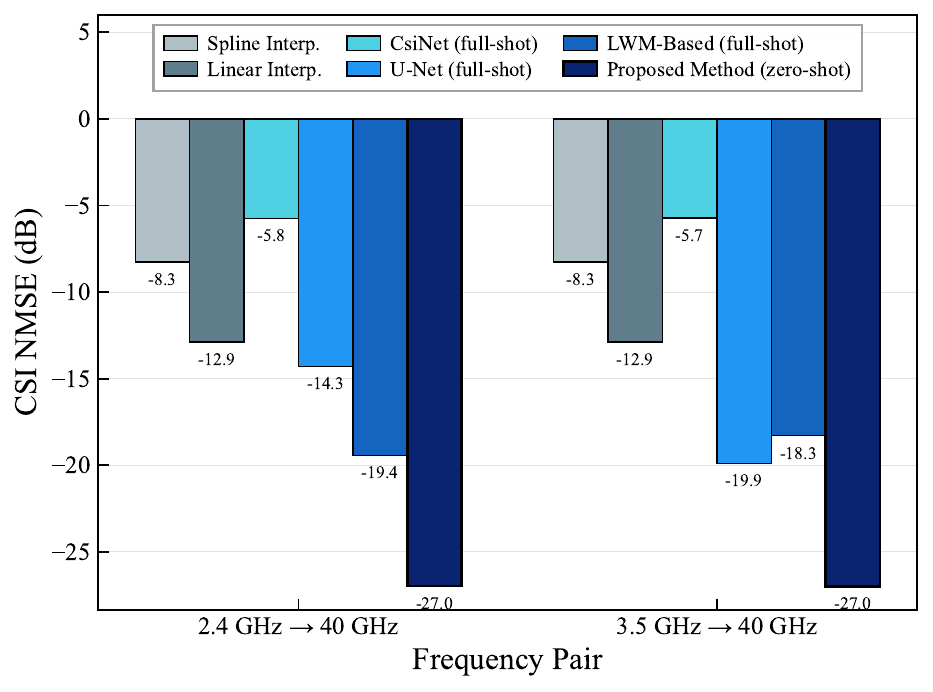}
\caption{Zero-shot transfer on unseen frequency pairs.}
\label{fig:zero_shot_results}
\end{figure}

Fig.~\ref{fig:zero_shot_results} evaluates whether the model can transfer
to unseen source--target frequency pairs. Our method is fine-tuned only on the six
supervised pairs in Table~\ref{tab:main_results} and is tested on
$2.4\rightarrow40$ and $3.5\rightarrow40$ GHz, with no paired samples from these
pairs used during fine-tuning. Although the baselines are trained full-shot on the
evaluated pairs, the proposed method still improves by $7.5$ and $7.1$~dB over the strongest
baseline, reaching $-27.0$~dB on both. Since $40$~GHz is seen during training only through
$15\rightarrow40$, the model recombines a known target condition with known source
conditions it never observed together rather than extrapolating to a 
new band.

\begin{figure}[!t]
\centering
\includegraphics[width=0.68\columnwidth]{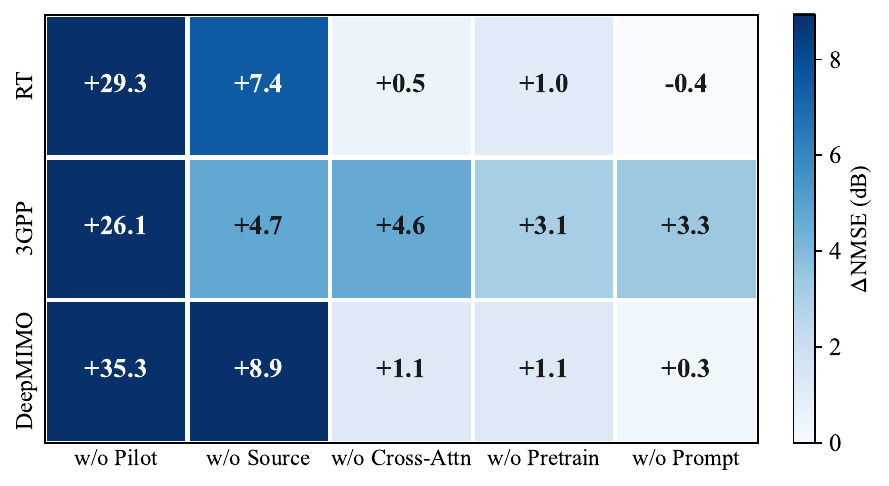}
\caption{Component ablation study.}
\label{fig:ablation_results}
\end{figure}

For ablation, Fig.~\ref{fig:ablation_results} removes one component at a time
under the same split, pilot mask, and SNR, and reports the degradation across
data sources. The two necessary inputs dominate everywhere: removing the target pilots costs
$26$--$35$~dB and removing the source CSI a further $4.7$--$8.9$~dB. Removing source CSI reduces the model to a pilot-only estimator, whose accuracy matches the dedicated U-Net baseline in Table~\ref{tab:main_results}, confirming the drop reflects the genuine value of cross-band information. The source
term helps most where the two bands share clean geometry (RT and
DeepMIMO city, $+7.4$ and $+8.9$~dB) and least on the stochastic 3GPP InH
channel. Cross-attention, RF prompting, and pretraining show the
opposite result: each is near-neutral on RT and DeepMIMO but contributes $3.1$--$4.6$~dB on 3GPP. Their small gains on RT and DeepMIMO do not indicate redundancy: on these channels pilots and source CSI already drive the model near a reconstruction ceiling, leaving little to remove, whereas on the hardest InH channel they recover $3.1$--$4.6$~dB. The modules are most helpful when the channel is more complex.

Table~\ref{tab:complexity} reports the inference cost on the $3.5\rightarrow28$ GHz
task in fp32 on a single RTX 4090 at batch size~1, where peak memory is the
maximum GPU memory allocated during one forward pass. The convolutional baselines
have the lowest latency but a clear accuracy tradeoff in
Table~\ref{tab:main_results}. Among the Transformer-scale models, the proposed method runs
faster than LWM and uses less peak memory despite its larger parameter count, as the compact
$8\times8$ token grid from the convolutional stem bounds attention cost and activation memory.

\begin{table}[!t]
\caption{Inference cost.}
\label{tab:complexity}
\centering
\scriptsize
\setlength{\tabcolsep}{3.5pt}
\renewcommand{\arraystretch}{1}

\begin{tabular}{@{}
>{\centering\arraybackslash}m{0.20\linewidth}
>{\centering\arraybackslash}m{0.14\linewidth}
>{\centering\arraybackslash}m{0.14\linewidth}
>{\centering\arraybackslash}m{0.16\linewidth}
>{\centering\arraybackslash}m{0.16\linewidth}
@{}}
\hline
\textbf{Method} & \textbf{Params (M)} & \textbf{FLOPs (G)} & \textbf{Latency (ms)} & \textbf{Peak Mem. (MB)} \\

\hline
CsiNet & 18.6 & 0.10 & 0.6 & 213 \\
U-Net & 1.9 & 2.31 & 1.4 & 464 \\
LWM & 3.1 & 5.45 & 10.7 & 6883 \\
Proposed Method & 40.7 & 6.26 & 9.6 & 505 \\
\hline
\end{tabular}
\end{table}

\section{Conclusion}
In this paper, we presented a foundation model for cross-band
CSI reconstruction with source-band CSI and sparse target-band pilots as inputs.
By aligning heterogeneous bands in a common PADS domain, fusing source and pilot
features, and using RF metadata as auxiliary conditioning, the model learns one
reconstruction function across multiple frequency pairs. Experiments show a
$6.1$~dB average improvement over the strongest supervised baseline and strong
zero-shot transfer to unseen source--target frequency pairs. These results suggest
that common PADS-based representations with
RF conditioning can reduce pair-specific adaptation while preserving practical inference latency and peak memory. Future work will
relax the requirement of synchronized source-band CSI and validate the approach
on measured multi-band channels.

\bibliography{references} 
\bibliographystyle{IEEEtran}

\end{document}